\documentstyle[12pt]{article}
\def\beq{\begin{equation}}
\def\eeq{\end{equation}}
\def\bea{\begin{eqnarray}}
\def\eea{\end{eqnarray}}
\def\bq{\begin{quote}}
\def\eq{\end{quote}}

\def\gappeq{\mathrel{\rlap {\raise.5ex\hbox{$>$}}
{\lower.5ex\hbox{$\sim$}}}}

\def\lappeq{\mathrel{\rlap{\raise.5ex\hbox{$<$}}
{\lower.5ex\hbox{$\sim$}}}}

\begin{document}
\topmargin -0.5cm
\oddsidemargin -0.3cm
\pagestyle{empty}
\vspace*{5mm}
\begin{center}
{\bf Multi-Hamiltonian Structure of Lotka-Volterra and Quantum Volterra Models}
\\
\vspace*{1.5cm}
{\bf Christofer Cronstr\"{o}m}$^{*)}$ \\
\vspace{0.3cm}
NORDITA \\
Blegdamsvej 17,\\
DK-2100 Copenhagen \O , Denmark\\
\vspace{0.5cm}
and\\
\vspace{0.5cm}
{\bf Milan Noga}$^{\dagger)}$ \\
\vspace{0.3cm}
Department of Theoretical Physics, Comenius University\\
Mlynska Dolina, Bratislava, Slovak Republic \\
\vspace*{3cm}
{\bf ABSTRACT} \\ \end{center}
\vspace*{5mm}
\noindent
We consider evolution equations of the Lotka-Volterra type, and elucidate
especially their
formulation as canonical Hamiltonian systems. The general conditions under
which these
equations admit several conserved quantities (multi-Hamiltonians) are analysed.
A special case,
which is related to the Liouville model on a lattice, is considered in detail,
both as a
classical and as a quantal system.

\vspace*{1cm}
\noindent

\noindent
$^{*)}$ Permanent address: Department of Theoretical Physics, University of
Helsinki, Finland.\\
e-mail: cronstrom@ phcu.helsinki.fi.\\ $^{\dagger)}$e-mail:
Milan.Noga@fmph.uniba.sk
\vspace*{1cm}
\vfill\eject

\setcounter{page}{1}
\pagestyle{plain}

\section{Introduction}  

The Lotka-Volterra model \cite{Lot,Goel} is defined by a set of non-linear
differential equations of the following form,
\begin{equation}
\frac{dw_{A}}{dt} = \epsilon_{A}w_{A} + \sum_{B=1}^{N} F_{AB}w_{A}w_{B}\;,\;A =
1,2,...N
\label{eq:LV1}
\end{equation}

In the equations above the quantities $w_{A}$, $A = 1,2,...,N$ are to be
determined in terms
of the given constant quantities $\epsilon_{A}$ and $F_{AB}$. The
Lotka-Volterra equations
are principally known as qualitative models in population dynamics with $N$
interacting species,
as well as rate equations for chemical reactions with $N$ constituents.
However,
the equations in question have actually turned out to be rather universal with
different
applications in physics \cite{Nov}, as witnessed e.g. by the case of the
Liouville model
formulated on a lattice \cite{Fadtak}.

The matrix $F$ in Eq.(\ref{eq:LV1}) ought to fulfill certain conditions which
relate
to the concept "crowding inhibits growth". This condition will here be  related
to the
requirement that the matrix $F$ be antisymmetric,
\begin{equation}
F_{AB} = - F_{BA}
\label{eq:antis}
\end{equation}

We wish to elucidate the circumstances under which the equations (\ref{eq:LV1})
including
the antisymmetry condition (\ref{eq:antis}), admit a canonical Hamiltonian
formulation.
This question has been answered in a
satisfactory manner by Kerner \cite{Kern} already quite some time ago, however
with the
restriction to an {\em even} number $N$ of species and with the condition that
the matrix
$F$ be {\em regular}. Here we give a general analysis, valid for even or odd
$N$, as well as for regular or singular matrices $F$. In the singular case
(which always
occurs if $N$ is odd), the equations (\ref{eq:LV1}) admit additional conserved
quantities
besides the Hamiltonian, under certain circumstances. We analyse these
circumstances in detail.
In special cases this phenomenon has been noted in the literature (e.g.
\cite{Nut}), and has
been referred to as a multi-Hamiltonian structure.

The analysis of the general Lotka-Volterra equations (\ref{eq:LV1}) including
the condition
(\ref{eq:antis}) is given in Sec. II below; the next Sec. III then deals with a
particular
example corresponding to an odd number $N = 3$ of species.

In Sec. IV we consider a specialisation of the Lotka-Volterra model, which
appeared in a
formulation of the Liouville model on the lattice \cite{Fadtak}, the quantal
formulation of
which was further considered by Volkov \cite{Volk} who introduced the name
{\em"Quantum Volterra
model"} in this connection.

The Hamiltonian proposed in \cite{Fadtak} and in \cite {Volk} is actually one
of the additional
conserved quantities that appear as a consequence of the special features of
the  model under
consideration, and differs from the Hamiltonian derived in this paper. Our
alternative
formulation permits a simple canonical quantization of the model under
consideration.

The final Sec. V contains a summary and discussion.

\vfill\eject

\section{The General Lotka-Volterra Equation}

In analysing Eq. (\ref{eq:LV1}) it is convenient to introduce new variables as
follows,
\begin{equation}
\xi_{A} = \log w_{A}
\label{eq:defxi}
\end{equation}
Then Eq. (\ref{eq:LV1}) can be rewritten as follows,
\begin{equation}
\dot{\xi}_{A} = \epsilon_{A} +  \sum_{B=1}^{N} F_{AB}\exp \xi_{B}
\label{eq:xi1}
\end{equation}
In order to analyse Eq. (\ref{eq:LV1}) further, it is convenient to refer to
the socalled
normal form of the antisymmetric matrix $F$.
It is well known \cite{Greub} that, by making an appropriate basis
transformation, any
antisymmetric $N\times N$ matrix $F$ can be transformed into the following
normal form:
\begin{eqnarray}
\pmatrix{0&-k_{1}&\ldots&\ldots&\ldots&\ldots&\ldots&\ldots\cr
+k_{1}&0&\ldots&\ldots&\ldots&\ldots&\ldots&\ldots\cr
\vdots&\vdots&\ddots&\vdots&\vdots&\vdots&\vdots&\vdots\cr
\ldots&\ldots&\ldots&0&-k_{n}&\ldots&\ldots&\ldots\cr
\ldots&\ldots&\ldots&+k_{n}&0&\ldots&\ldots&\ldots\cr
\ldots&\ldots&\ldots&\ldots&\ldots&0&\ldots&\ldots\cr
\ldots&\ldots&\ldots&\ldots&\ldots&\ldots&\ddots&\ldots\cr
\ldots&\ldots&\ldots&\ldots&\ldots&\ldots&\ldots&0\cr}
\label{eq:Narr}
\end{eqnarray}
where the (positive) quantities $k_{\alpha},\; \alpha = 1,...,n$ are the square
roots
of those characteristic values of the matrix $- F^{2}$, which are different
from zero.
The number $n$ of non-zero characteristic values $k_{\alpha}^{2}\;$is thus
given
by the rank ($2n$, say) of the matrix $F$.

It is furthermore convenient to relate the normal form (\ref{eq:Narr})
to the following linear equations,
\begin{equation}
\sum_{B=1}^{N} F_{AB} x_{\alpha B} = +\; k_{\alpha} y_{\alpha A},\; \alpha =
1,...,n
\label{eq:ex}
\end{equation}
\begin{equation}
\sum_{B=1}^{N} F_{AB} y_{\alpha B} = -\; k_{\alpha} x_{\alpha A},\; \alpha =
1,...,n
\label{eq:yle}
\end{equation}
and
\begin{equation}
\sum_{B=1}^{N} F_{AB} z_{\beta B} = 0,\; \beta = 1,...,N-2n
\label{eq:zed}
\end{equation}
with the understanding that Eq. (\ref{eq:zed}) is empty if the rank of the
matrix
$F$ is $N$ (so that $N = 2n$), in which case  the  matrix $F$ is {\em regular},
\begin{equation}
det\; F \neq 0.
\label{eq:regN}
\end{equation}

The cases $N$ even or odd differ qualitatively in general, since in the latter
case the
matrix $F$ is {\em necessarily} singular.  Then  Eq. (\ref{eq:zed}) has an odd
number
of nontrivial solutions. Eq. (\ref{eq:zed}) may of course also have non-trivial
solutions if
$N$ is even, in which case there is necessarily an even number of such
solutions.

We ortho-normalize the vectors $x_{\alpha A}$, $y_{\alpha A}$, and $z_{\beta
A}$,  properly,
\begin{equation}
(x_{\alpha}, x_{\beta}) = \delta_{\alpha \beta},\: (y_{\alpha}, y_{\beta}) =
\delta_{\alpha \beta},\:(z_{\alpha}, z_{\beta}) = \delta_{\alpha \beta}
\label{eq:norma}
\end{equation}
where the inner product $(u, v)$ of any vectorlike quantities $u_{A}$ and
$v_{A}$
is defined as follows,
\begin{equation}
(u, v) \equiv \sum_{A=1}^{N} u_{A} v_{A}
\label{eq:innerp}
\end{equation}
The equations (\ref{eq:ex}), (\ref{eq:yle}) and (\ref{eq:zed}) imply the
following orthogonality relations,
\begin{equation}
(x_{\alpha}, y_{\beta}) = 0,\: \alpha, \beta = 1,...,n;
\label{eq:xyorth}
\end{equation}
and
\begin{equation}
(x_{\alpha}, z_{\beta}) = (y_{\alpha}, z_{\beta}) = 0, \:\alpha = 1,...,n,
\beta = 1,...,N-2n
\label{eq:zorth}
\end{equation}

After these preliminaries we return to the equations (\ref{eq:LV1}).
Contracting Eq.
(\ref{eq:LV1}) with any solution $z_{\beta}$ of Eq. (\ref{eq:zed}), one
obtains, in view
of the antisymmetry condition (\ref{eq:antis}),
\begin{equation}
(z_{\beta}, \dot{\xi}) = (z_{\beta}, \epsilon) \equiv r_{\beta}
\label{eq:zedxi}
\end{equation}
where we use the inner product notation given above in Eq. (\ref{eq:innerp}).
Thus,
\begin{equation}
(z_{\beta}, \xi) = r_{\beta}t + K_{\beta}\;,\; \beta = 1,...,N-2n
\label{eq:consxi}
\end{equation}
where the quantities $K_{\beta}$ are arbitrary constants of integration. The
equations
(\ref{eq:consxi}) define $N-2n$ {\em constraints} among the variables
$\xi_{A}$; the number
of these constraints is equal to the number of linearly independent solutions
(if any) to the
zero-eigenvalue equations (\ref{eq:zed}). In particular, if
\begin{equation}
r_{\beta} \equiv (z_{\beta}, \epsilon) \neq 0\; \beta = 1,...,N-2n
\label{eq:zeps}
\end{equation}
then the $N-2n$ quantities $(z_{\beta}, \xi)$ are linear functions of $t$.

The equations (\ref{eq:LV1}) still contain $2n$ unknowns; one obtains a
convenient
set of $2n$ equations for these unknowns by contracting the equations
(\ref{eq:LV1})
with the solutions $x_{\alpha}$ and $y_{\alpha}$ of the equations (\ref{eq:ex})
and
(\ref{eq:yle}), respectively. Thus,
\begin{equation}
(x_{\alpha}, \dot{\xi}) =
(x_{\alpha}, \epsilon) - \sum_{B=1}^{N} k_{\alpha} y_{\alpha B} \exp \xi_{B}
\label{eq:exxi}
\end{equation}
and
\begin{equation}
(y_{\alpha}, \dot{\xi}) =
(y_{\alpha}, \epsilon) + \sum_{B=1}^{N} k_{\alpha} x_{\alpha B} \exp \xi_{B}
\label{eq:yxi}
\end{equation}

It should be noted that the equations (\ref{eq:zedxi}), (\ref{eq:exxi}) and
(\ref{eq:yxi})
are equivalent to the original equations (\ref{eq:LV1}) under the condition
(\ref{eq:antis}),
since the orthonormal vectors $x_{\alpha}$, $y_{\alpha}$ and $z_{\alpha}$ form
a complete set
in $N$ dimensions. Thus,
\begin{equation}
\xi_{A} = \sum_{\alpha =1}^{n} \left [ (x_{\alpha}, \xi) x_{\alpha A}
+ (y_{\alpha}, \xi) y_{\alpha A} \right ] + \sum_{\beta = 1}^{N-2n} (z_{\beta},
\xi) z_{\beta A}
\label{eq:compl}
\end{equation}

The {\em even} number of equations (\ref{eq:exxi}) and (\ref{eq:yxi}) are in
fact Hamiltonian
equations in coordinates $q_{\alpha}$ and momenta $p_{\alpha}$ defined as
follows,
\begin{equation}
p_{\alpha} \equiv C_{\alpha} (x_{\alpha}, \xi)
\label{eq:canp}
\end{equation}
and
\begin{equation}
q_{\alpha} \equiv D_{\alpha} (y_{\alpha}, \xi)
\label{eq:canq}
\end{equation}
where the quantities $C_{\alpha}$ and $D_{\alpha}$ are constants, which satisfy
the following
condition,
\begin{equation}
C_{\alpha}D_{\alpha}k_{\alpha} = 1
\label{eq:CDk}
\end{equation}
Namely, let
\begin{equation}
H(p,q;t) \equiv \sum_{B=1}^{N} \exp \left \{\xi_{B}(p, q; t) \right \} -
\sum_{\alpha =1}^{n}
 \left [C_{\alpha}
(x_{\alpha}, \epsilon) q_{\alpha} - D_{\alpha}(y_{\alpha}, \epsilon) p_{\alpha}
\right ]
\label{eq:Cham}
\end{equation}
where the quantity $\xi(p, q; t)$ is expressed in terms of the quantities
$p_{\alpha}$,
$q_{\alpha}$ defined above, as well as in terms of the quantities $r_{\beta}$
and $K_{\beta}$
defined in Eq.
(\ref{eq:zedxi}) and (\ref{eq:consxi}), respectively,  according to Eq.
(\ref{eq:compl}),
\begin{equation}
\xi_{A}(p, q; t) = \sum_{\alpha =1}^{n} \left [ (C_{\alpha}^{-1} p_{\alpha})
x_{\alpha A} +
(D_{\alpha}^{-1} q_{\alpha}) y_{\alpha A} \right ] + \sum_{\beta =1}^{N-2n}
(r_{\beta}t +
K_{\beta})z_{\beta A}
\label{eq:xpqr}
\end{equation}
Using Eq. (\ref{eq:xpqr}) and Eq. (\ref{eq:CDk}) one obtains straightforwardly
the following
results from Eq. (\ref{eq:Cham}),
\begin{equation}
\frac{\partial H}{\partial p_{\alpha}} = D_{\alpha} \left [
\sum_{B=1}^{N}k_{\alpha}x_{\alpha B}
\exp \left \{\xi_{B}(p, q; t) \right \} + (y_{\alpha}, \epsilon) \right ]
\label{eq:pDHp}
\end{equation}
and
\begin{equation}
\frac{\partial H}{\partial q_{\alpha}} = C_{\alpha} \left [
\sum_{B=1}^{N}k_{\alpha}y_{\alpha B}
\exp \left \{\xi_{B}(p, q; t) \right \} - (x_{\alpha}, \epsilon) \right ]
\label{eq:pDHq}
\end{equation}
Then, using the relations just given, it is a simple matter to verify that Eq.
(\ref{eq:exxi})
is nothing but the following,
\begin{equation}
\frac{dp_{\alpha}}{dt} = - \; \frac{\partial H}{\partial q_{\alpha}}
\label{eq:cpdot}
\end{equation}
Likewise, Eq. (\ref{eq:yxi}) is equivalent to the following,
\begin{equation}
\frac{dq_{\alpha}}{dt} = + \; \frac{\partial H}{\partial p_{\alpha}}
\label{eq:cqdot}
\end{equation}

It should be noted that the Hamiltonian (\ref{eq:Cham}) is in general {\em
explicitly
time-dependent} if the zero-eigenvalue equations (\ref{eq:zed}) have
non-trivial solutions.
This requires that the
matrix $F$ ocurring in the Lotka-Volterra equations (\ref{eq:LV1}) be {\em
singular},
which is always the case if the dimensionality $N$ of the system is an odd
integer. Be that as
it may, as we have just demonstrated, there is in any case always an {\em
even-dimensional}
subset of the equations
(\ref{eq:LV1}) which can be written in canonical Hamiltonian form, in terms of
canonical
momenta (\ref{eq:canp}) and coordinates (\ref{eq:canq}), with the expression
(\ref{eq:Cham})
as a Hamiltonian.

The explicit time-dependence of the Hamiltonian (\ref{eq:Cham}) disappears if
the vector
$\epsilon$ (the array of rate-constants $\epsilon_{A}$) occuring in Eq.
(\ref{eq:LV1}) is
orthogonal to all the solutions $z_{\beta}$ of the zero-eigenvalue equation
(\ref{eq:zed}),
\begin{equation}
(z_{\beta}, \epsilon) = 0\;, \; \beta = 1,...,N-2n
\label{eq:zednul}
\end{equation}
If the equations (\ref{eq:zednul}) are in force,  we have, according to Eq.
(\ref{eq:consxi}),
\begin{equation}
(z_{\beta}, \xi) =   K_{\beta}\;,\; \beta = 1,...,N-2n
\label{eq:zedcon}
\end{equation}

Each orthogonality condition (\ref{eq:zednul}) with a fixed value of the index
$\beta$,
gives rise to a {\em conserved} quantity, namely the corresponding expression
in
Eq. (\ref{eq:zedcon}), according to Eqns. (\ref{eq:consxi}).

We conclude this section by summarizing the results obtained so far:

The Lotka-Volterra equations (\ref{eq:LV1}), including the condition
(\ref{eq:antis}), can
always be written as a system of canonical Hamiltonian equations in an even
number of
unconstrained canonical variables, together with a set of explicitly solvable
constraints,
which are time-dependent in general. This set is non-empty, i.e. the
constraints in question
occur in general, if the matrix $F$  in Eq. (\ref{eq:LV1}) is singular. The
number of constraints equals the number of linearly independent eigen-vectors
$z_{\beta}$ of
the matrix $F$, which correspond to zero-eigenvalues.

For non-singular matrices $F$ the Hamiltonian does not depend on time
explicitly. In the
singular case, the Hamiltonian is in general explicitly time-dependent. This
time-dependence
disappears from the Hamiltonian, if all the eigen-vectors $z_{\beta}$ are
orthogonal to the
vector $(\epsilon_{1},...,\epsilon_{N})$ made up by the rate-constants
$\epsilon_{A}$
in the basic equations (\ref{eq:LV1}). Each orthogonality condition of the
aforementioned kind
gives rise to a (time-independent) linear constraint among the variables
$\xi_{A}$ in the
equations (\ref{eq:xi1}), which are equivalent to the basic equations
(\ref{eq:LV1}).

\vfill\eject

\section{An Example Involving Three Degrees of Freedom}

We consider below a system of the Lotka-Volterra type, with three degrees of
freedom, which has
been analysed in great detail by Grammaticos et.al \cite{Gram}, with the
objective of making a
systematic search for first integrals of the system in question.
A special case of this system has been presented by Nutku \cite {Nut} as an
example of a
bi-Hamiltonian system of the Lotka-Volterra type.

The general system is the follwing ($a$, $b$, $c$ and $\lambda$, $\mu$, $\nu$
are constants),
\begin{equation}
\dot{s} = s(\lambda + ct + u)\;, \dot{t} = t(\mu + s + au)\;, \dot{u} = u(\nu +
bs + t)
\label{eq:Gramex}
\end{equation}
Rescaling the variables in Eq. ({\ref{eq:Gramex}) as follows,
\begin{equation}
s \rightarrow \frac{1}{\alpha_{1}}s \equiv w_{1}\;,t \rightarrow
\frac{1}{\alpha_{2}}t
\equiv w_{2}\; u \rightarrow \frac{1}{\alpha_{3}}u \equiv w_{3}
\label{eq:scale}
\end{equation}
where the parameters $\alpha_{n}, n=1,2,3,$ are at our disposal, and using the
$\xi$-variables
introduced in (\ref{eq:defxi}), we get an equation of the form (\ref{eq:xi1})
from Eq.
(\ref{eq:Gramex}),
\begin{equation}
\dot{\xi}_{A} = \epsilon_{A} + \sum_{B=1}^{3} F_{AB} \exp \xi_{B}
\label{eq:Gekko}
\end{equation}
where,
\begin{equation}
\epsilon_{1} =  \lambda\;,\,\epsilon_{2} = \mu\;,\;\epsilon_{3} =\nu
\label{eq:Graeps}
\end{equation}
and
\begin{eqnarray}
F =
\pmatrix{0&\alpha_{2}c&\alpha_{3}\cr
\alpha_{1}&0&\alpha_{3}a\cr
\alpha_{1}b&\alpha_{2}&0\cr}
\label{eq:GramN}
\end{eqnarray}
The matrix $F$ defined by Eq. (\ref{eq:GramN}) does not fulfill the
antisymmetry
condition (\ref{eq:antis}) as such, but becomes antisymmetric under the
follwing conditions,
\begin{equation}
\alpha_{1} = - \alpha_{2}c\;, \alpha_{2} = - \alpha_{3}a\;, \alpha_{3} = -
\alpha_{1}b
\label{eq:alfanti}
\end{equation}
But Eqs. (\ref{eq:alfanti}) can be fulfilled if and only if
\begin{equation}
abc = -1
\label{eq:abc}
\end{equation}
{}From now on, we {\em assume} that the parameters $a, b, c$ satisfy the
condition (\ref{eq:abc}).
Without essential loss of generality, we can choose one parameter $\alpha_{n}$
at will in
the equations above. Taking $\alpha_{1} = 1$, we satisfy the equations
(\ref{eq:alfanti}) as
follows,
\begin{equation}
\alpha_{1} = 1, \;\alpha_{2} = ab, \; \alpha_{3} = - \;b
\label{eq:final}
\end{equation}

We then apply the results of Sec. II to the system of equations above.

In the first place we have to consider the following system of linear
(eigenvalue) equations
(compare (\ref{eq:ex}), (\ref{eq:yle}) and (\ref{eq:zed})),
\begin{equation}
\sum_{B=1}^{3} F_{AB}x_{B} = +\;ky_{A}\;,\; \sum_{B=1}^{3} F_{AB}y_{B} = -\;k
x_{A}\;,\;
\sum_{B=1}^{3} F_{AB} z_{B} = 0
\label{eq:eigen}
\end{equation}
The following are appropriately orthonormalised solutions to these equations,
\begin{eqnarray}
x = \pmatrix {0\cr\frac{1}{\sqrt{1+b^{2}}}\cr\frac{b}{\sqrt{1+b^{2}}}\cr}\;,\;
y = \frac{1}{k}\pmatrix {- \sqrt{1+b^{2}}\cr-\frac{ab^{2}}{\sqrt{1+b^{2}}}\cr
\frac{ab}{\sqrt{1+b^{2}}}\cr}\;,\;
z = \frac{1}{k}\pmatrix {ab\cr-b\cr1\cr}
\label{eq:colxyz}
\end{eqnarray}
with
\begin{equation}
k = \sqrt{1+b^{2} + a^{2}b^{2}}
\label{eq:kay}
\end{equation}
Using Eqns. (\ref{eq:colxyz}) one now obtains the follwing constraint,
according to Eq.
(\ref{eq:consxi}),
\begin{equation}
(z, {\xi}) \equiv \frac{1}{k} [ab\xi_{1} - b\xi_{2} + \xi_{3}] = r_{1}t + K_{1}
\label{eq:Gzedxi}
\end{equation}
where
\begin{equation}
r_{1} \equiv (z, \epsilon) = \frac{1}{k} [ab\lambda - b\mu + \nu]
\label{eq:Gerr}
\end{equation}
The remaining variables can then be taken to be the canonical variables defined
in Eqns.
(\ref{eq:canp}, \ref{eq:canq}), i.e. in the present case,
\begin{equation}
p \equiv C(x, \xi) = \frac{C}{\sqrt{1+b^{2}}} (\xi_{2} + b\xi_{3})
\label{eq:Gpe}
\end{equation}
and
\begin{equation}
q \equiv D(y, \xi) = -\; \frac{D}{k\sqrt{1+b^{2}}}( (1+b^{2})\xi_{1} +
ab^{2}\xi_{2} -
ab\xi_{3})
\label{eq:Gqu}
\end{equation}
where the constants $C$ and $D$ are related as follows
\begin{equation}
CDk = 1
\label{eq:GCDk}
\end{equation}
but are otherwise arbitrary.

The equations (\ref{eq:Gzedxi}), (\ref{eq:Gpe}) and (\ref{eq:Gqu}) can easily
be inverted,

\begin{eqnarray}
\xi_{1} & = &\frac{ab}{k}(r_{1}t+K_{1}) - \sqrt{1+b^{2}} Cq \nonumber \\
\xi_{2} & = &- \frac{b}{k}(r_{1}t+K_{1}) + \frac{k}{\sqrt{1+b^{2}}} Dp -
\frac{ab^{2}}{\sqrt{1+b^{2}}} Cq \\
\xi_{3} & = &\frac{1}{k}(r_{1}t+K_{1}) + \frac{bk}{\sqrt{1+b^{2}}} Dp +
\frac{ab}{\sqrt{1+b^{2}}} Cq \nonumber
\label{eq:eks123}
\end{eqnarray}

Using the results above, it is simple to verify that the variables $p$ and $q$
defined
in the equations (\ref{eq:Gpe}), (\ref{eq:Gqu}) above, satisfy canonical
Hamiltonian equations,
i.e. the equations (\ref{eq:cpdot}), (\ref{eq:cqdot}), with the following
Hamiltonian $H$,
\begin{equation}
H =   \frac{1}{k^{2}} \left [ (\mu + b\nu)\xi_{1} - (\lambda - ab\nu)\xi_{2}
- b(\lambda + a\mu)\xi_{3} \right ] + \sum_{A=1}^{3} \exp \xi_{A}
\label{eq:GekkoH}
\end{equation}

The Hamiltonian (\ref{eq:GekkoH}) is in general explicitly {\em time
dependent}, since the
sum of exponentials in (\ref{eq:GekkoH}) above, depends on time in general, as
can be inferred
from the equations (\ref{eq:Gzedxi}). However, if the orthogonality condition
(\ref{eq:zednul})
is in force, i.e. if
\begin{equation}
r_{1} \equiv (z, \epsilon) = \frac{1}{k} [ab\lambda - b\mu + \nu] = 0
\label{eq:Gorthog}
\end{equation}
then the Hamiltonian (\ref{eq:GekkoH}) becomes a constant of motion. However,
if the condition
(\ref{eq:Gorthog}) is in force, then there exists a {\em second} constant of
motion, according
to Eq.(\ref{eq:Gzedxi}), namely the following,
\begin{equation}
\frac{1}{k} [ab\xi_{1} - b\xi_{2} + \xi_{3}]
\label{eq:GH2}
\end{equation}

The appearance of the two conserved quantities (\ref{eq:GekkoH}) and
(\ref{eq:GH2}) under
the condition (\ref{eq:Gorthog}) is just the phenomenon which has been referred
to by the term
bi-Hamiltonian in the present context by Nutku \cite{Nut}.

This terminology is perhaps a little unfortunate, despite the fact that the
quantity
(\ref{eq:GH2}) can be understood as a Hamiltonian for the system
(\ref{eq:Gekko}) (if condition
(\ref{eq:Gorthog}) is in force) in a certain general sense \cite{Olver}.
On the contrary, the
Hamiltonian (\ref{eq:GekkoH}), is a Hamiltonian in a strict sense, regardless
of whether
the condition (\ref{eq:Gorthog}) is true or not. By strict sense is here meant
that the
formulation involving the Hamiltonian (\ref{eq:GekkoH}) is explicitly
canonical, involving
a known pair of canonical variables $p$ and $q$. So, in the general case, only
the quantity
(\ref{eq:GekkoH}) merits a designation as a Hamiltonian in the strict canonical
sense employed
in this paper.

The considerations above are naturally generalisable to systems with more than
three degrees of
freedom, in view of the general results given in Sec. II.

\vfill\eject

\section{The Liouville Model on a Lattice}

The Liouville model on a lattice discussed by Faddeev and Takhtajan
\cite{Fadtak} and by Volkov
\cite{Volk} is specified by the following classical equations of motion, which
is a special system of
Lotka-Volterra equations,
\begin{equation}
\frac{dw_{A}}{dt} =  \frac{1}{2\Delta } w_{A}(w_{A+1} - w_{A-1})\;,\; A =
1,...,N
\label{eq:QV1}
\end{equation}
where $N$ is an {\em even} positive integer, and where $\Delta$ is a parameter.
The variables $w_{A}$ are furthermore supposed to satisfy the following
periodicity condition,
\begin{equation}
w_{N+n} = w_{n}\;, \; n = 0,1,2,...
\label{eq:percond}
\end{equation}
Comparing Eqs. (\ref{eq:QV1}) with the general Lotka-Volterra equations
(\ref{eq:LV1})
one observes that the  model defined above by the Eqns. (\ref{eq:QV1})
corresponds to the case
of vanishing rate constants $\epsilon_{A}$,
\begin{equation}
\epsilon_{A} = 0\;,\; A = 1,2,...,N
\label{eq:vaneps}
\end{equation}
and to the case in which the matrix $F$ has the following form,
\begin{eqnarray}
(F_{AB}) =
\pmatrix{0&\lambda&\ldots&\ldots&\ldots&\ldots&\ldots&-\lambda\cr
-\lambda&0&\lambda&\ldots&\ldots&\ldots&\ldots&\ldots\cr
\vdots&\ddots&\ddots&\ddots&\vdots&\vdots&\vdots&\vdots\cr
\ldots&\ldots&-\lambda&0&\lambda&\ldots&\ldots&\ldots\cr
\ldots&\ldots&\ldots&-\lambda&0&\lambda&\ldots&\ldots\cr
\ldots&\ldots&\ldots&\ldots&\ddots&\ddots&\ddots&\ldots\cr
\ldots&\ldots&\ldots&\ldots&\ldots&-\lambda&0&\lambda\cr
\lambda&\ldots&\ldots&\ldots&\ldots&\ldots&-\lambda&0\cr}
\label{eq:QVN}
\end{eqnarray}
where
\begin{equation}
\lambda \equiv \frac{1}{2\Delta}
\label{eq:lamb}
\end{equation}
The matrix (\ref{eq:QVN}) is {\em singular},
\begin{equation}
det \; F = 0
\label{eq:QNdet}
\end{equation}
This circumstance together with the conditions (\ref{eq:vaneps}) implies the
existence
of additional conserved quantities (constants of motion) besides the
Hamiltonian,
according to the general results presented in Sec. II.

At first we give a brief analysis of the canonical formulation for the
classical latticised
Liouville model, along the lines presented in Sec. II.

\vfill\eject

\subsection{Canonical Hamiltonian Formulation of the \newline
Classical Liouville Model on a Lattice}

First we rewrite Eq. (\ref{eq:QV1}) using the variable (\ref{eq:defxi}),
\begin{equation}
\dot{\xi}_{A} = \lambda \left ( \exp \xi_{A+1} - \exp \xi_{A-1} \right ) \equiv
\sum_{B=1}^{N} F_{AB} \exp \xi_{B}
\label{eq:Qxi1}
\end{equation}
where the matrix $F$ given by the expression (\ref{eq:QVN}).
According to the general method developed in Sec. II, we consider  to begin
with the
eigenvalue-equation (\ref{eq:zed}), i.e. the following equation,
\begin{equation}
\sum_{B=1}^{N} F_{AB} z_{\beta, B} = 0
\label{eq:Qzed}
\end{equation}

As is immediately verified, Eq. (\ref{eq:Qzed}) has two linearly independent
(normalized)
solutions $z_{\beta}$, which we take to be the following,
\begin{equation}
z_{1,A} =  \frac{1}{\sqrt{2N}}(1 - (-1)^{A})\;,\;  A = 1,...,N
\label{eq:Qzed1}
\end{equation}
and
\begin{equation}
z_{2,A} = \frac{1}{\sqrt{2N}}(1 + (-1)^{A}) \;,\;  A = 1,...,N
\label{eq:Qzed2}
\end{equation}

The existence of the two non-trivial solutions (\ref{eq:Qzed1}) and
(\ref{eq:Qzed2})
implies the existence of the following two conserved quantities, according to
Eqs.
(\ref{eq:zednul}) and (\ref{eq:zedcon}),
\begin{equation}
\sqrt{\frac{2}{N}} \sum_{A=1}^{\frac{1}{2}N} \xi_{2A-1} = K_{1}
\label{eq:firstc}
\end{equation}
and
\begin{equation}
\sqrt{\frac{2}{N}} \sum_{A=1}^{\frac{1}{2}N} \xi_{2A} = K_{2}
\label{eq:secondc}
\end{equation}
Needless to say, the Hamiltonian (\ref{eq:Cham}), which now takes the following
simple form, is also conserved,
\begin{equation}
H = \sum_{B=1}^{N} \exp \xi_{B} \equiv \sum_{B=1}^{N} w_{B}
\label{eq:QCham}
\end{equation}

We then have to consider the equations (\ref{eq:ex}) and (\ref{eq:yle}) with
the matrix
$F$ given by (\ref{eq:QVN}).

After calculations one finds the follwing solutions $x_{\alpha A}$ and
$y_{\alpha A}$,
\begin{equation}
x_{\alpha A} = \frac{1}{\sqrt{2N}}(1 + (-1)^{A})\left [\cos (A\varphi_{\alpha})
+ \sin
(A\varphi_{\alpha}) \right ]\;, \;\alpha = 1,2,...,\frac{1}{2}N -1.
\label{eq:Qxalfa}
\end{equation}
and
\begin{equation}
y_{\alpha A} = \frac{1}{\sqrt{2N}}(1 - (-1)^{A})\left [\cos (A\varphi_{\alpha})
- \sin
(A\varphi_{\alpha}) \right ]\;, \;\alpha = 1,2,...,\frac{1}{2}N -1.
\label{eq:Qyalfa}
\end{equation}
where
\begin{equation}
\varphi_{\alpha} = \frac{2\pi \alpha}{N}\;,\;\alpha = 1,2,...,\frac{1}{2}N -1.
\label{eq:varphi}
\end{equation}
The eigenvalue parameter $k_{\alpha}$ is expressed in terms of
$\varphi_{\alpha}$ as
follows,
\begin{equation}
k_{\alpha} = 2\lambda \sin \varphi_{\alpha}\;,\;\alpha = 1,2,...,\frac{1}{2}N
-1.
\label{eq:kalfa}
\end{equation}

Applying the general formulae (\ref{eq:canp}) and (\ref{eq:canq}) we now obtain
the
following expressions for the canonical momenta and coordinates, respectively,
\begin{equation}
p_{\alpha} =  \frac{2C_{\alpha}}{\sqrt{N}} \sum_{A=1}^{\frac{1}{2}N}
 \xi_{2A} \sin [2A\varphi_{\alpha} + \frac{\pi}{4}]
\label{eq:CCp}
\end{equation}
and
\begin{equation}
q_{\alpha} =  \frac{2D_{\alpha}}{\sqrt{N}} \sum_{A=1}^{\frac{1}{2}N}
\xi_{2A-1}\cos [(2A-1)\varphi_{\alpha} + \frac{\pi}{4}]
\label{eq:CCq}
\end{equation}

The inverses of Eqns. (\ref{eq:CCp}), (\ref{eq:CCq}) are obtained
straightforwardly, either
by using the general result (\ref{eq:xpqr}) or by evaluating the appropriate
sums involving
trigonometric functions,
\begin{equation}
\xi_{A} = \sum_{\alpha} C^{-1}_{\alpha}p_{\alpha} x_{\alpha A} +
\sum_{\alpha} D^{-1}_{\alpha}q_{\alpha} y_{\alpha A} + \sum_{\beta=1}^{2}
K_{\beta}z_{\beta, A}
\label{eq:CCxipqK}
\end{equation}
where the constants $K_{\beta}$ are given in Eqns. (\ref{eq:firstc}),
(\ref{eq:secondc}).

We conclude this sub-section by giving a Poisson-bracket formulation of the
equations
(\ref{eq:Qxi1}). The basic Poisson-bracket is the following
\begin{equation}
\left \{ \xi_{A}, \xi_{B} \right \}_{PB} \equiv
\sum_{\alpha =1}^{\frac{1}{2}N-1}\left [ \frac{\partial \xi_{A}}{\partial
q_{\alpha}}
\frac{\partial \xi_{B}}{\partial p_{\alpha}} - \frac{\partial \xi_{B}}{\partial
q_{\alpha}}
\frac{\partial \xi_{A}}{\partial p_{\alpha}}\right ] = - \; \sum_{\alpha
=1}^{n}
k_{\alpha}\left [ x_{\alpha A}y_{\alpha B} - y_{\alpha A}x_{\alpha B} \right ]
\label{eq:xiPB}
\end{equation}
Finally, evaluating the
relevant trigonometric sums in (\ref{eq:xiPB}) (compare Eqns.
(\ref{eq:Qxalfa}),
(\ref{eq:Qyalfa})), one obtains,
\begin{equation}
\left \{ \xi_{A}, \xi_{B} \right \}_{PB} = \lambda \left ( \delta_{A,B-1} -
\delta_{A,B+1}
\right ) \equiv F_{AB}
\label{eq:CxiPB}
\end{equation}

Using the result (\ref{eq:CxiPB}) it is straightforward to verify that the
equations
(\ref{eq:Qxi1}) are reproduced by the following expressions,
\begin{equation}
\dot{\xi}_{A} = \left \{ \xi_{A}, H \right \}_{PB} =
\sum_{C=1}^{N} \left \{ \xi_{A}, \xi_{C} \right \}_{PB} \frac{\partial
H}{\partial \xi_{C}}
\label{eq:PBQxi1}
\end{equation}
where the Hamiltonian is given by Eq. (\ref{eq:QCham})

Reverting to the original variables $w_{A}$  (compare Eq. (\ref{eq:defxi})),
one gets the
following Poisson-bracket relations from (\ref{eq:CxiPB}),
\begin{equation}
\left \{ w_{A}, w_{B} \right \}_{PB} = \lambda w_{A}w_{B}\left ( \delta_{A,B-1}
- \delta_{A,B+1} \right )
\label{eq:CwPB}
\end{equation}

It is perhaps necessary to emphasize that the result (\ref{eq:CwPB}) is firmly
based on the
canonical structure derived above for the latticised Liouville model, and not
merely an
example of a bracket structure which is consistent with the Hamiltonian
(\ref{eq:QCham})
and the equations of motion (\ref{eq:QV1}).

In the aforementioned paper by Volkov \cite{Volk}, an alternative Hamiltonian
formalism
to the one presented above has been proposed for the quantization of the system
defined by the equations (\ref{eq:QV1}).

The Hamiltonian $H_{V}$ used by Volkov is simply the sum of the two constants
of motion
(\ref{eq:firstc}) and (\ref{eq:secondc}), apart from a multiplicative constant,
\begin{equation}
H_{V} = - \frac{1}{2\gamma \Delta} \sum_{A=1}^{N} \log w_{A}
\label{eq:Volham}
\end{equation}
where $\gamma$ is an arbitrary coupling constant, which is introduced for
convenience.
The quantity (\ref{eq:Volham}) is a constant of motion, since it is the sum of
two constants
of motion (compare Eqns. (\ref{eq:firstc}, \ref{eq:secondc})).

The Hamiltonian scheme of Volkov consists of a bracket formulation of the
equations
(\ref{eq:QV1}) with (\ref{eq:Volham}) as a Hamiltonian. Thus,
\begin{equation}
\dot{w}_{A} = \left \{w_{A},H_{V}\right \} = \sum_{B=1}^{N}\left
\{w_{A},w_{B}\right \}
\frac{\partial H_{V}}{\partial w_{B}}
\label{eq:Volbreq}
\end{equation}
The basic bracket given by Volkov is the following (after correction of a
crucial printing
error; an overall factor $w_{A}w_{B}$ is missing from Volkovs expression, Eq.
(23) of
Ref. \cite{Volk}),
\begin{equation}
\left \{w_{A}, w_{B} \right \} = \frac{1}{2} \gamma w_{A}w_{B} \left [ (4 -
w_{A} - w_{B})
(\delta_{A+1,B} - \delta_{A-1,B}) + w_{A-1}\delta_{A-2,B}
-w_{A+1}\delta_{A+2,B} \right ]
\label{eq:Vbracket}
\end{equation}

It is indeed simple to verify, that the equations (\ref{eq:QV1}) are reproduced
as bracket
equations (\ref{eq:Volbreq}) if one uses the bracket (\ref{eq:Vbracket}).
However, it should
be noted that the Hamiltonian $H_{V}$ defined in Eq. (\ref{eq:Volham}) which is
used in this
context, is not bounded from
below, and does therefore not permit the construction of classical statistical
mechanics for the
(classical) system in question. On the other hand, the Hamiltonian $H$ defined
in
Eq. (\ref{eq:QCham}), which has been obtained here on the basis of general
considerations
of the canonical structure of systems of the Lotka-Volterra type, {\em is}
bounded from below,
and is thus preferable also from this point of view.

\vfill\eject

\subsection{The Canonical Quantization of the \newline Liouville Model on a
Lattice}

The quantization of the model considered above is straightforward if one uses
the canonical
structure derived here as a starting point. Following the usual prescription of
replacing
Poisson brackets involving classical canonical variables by commutators
involving the
corresponding operators one has the following canonical quantum operator
relations,
\begin{equation}
[\hat{p}_{\alpha}, \hat{p}_{\beta}] = 0, \;[\hat{q}_{\alpha}, \hat{q}_{\beta}]
= 0,\;
[\hat{q}_{\alpha}, \hat{p}_{\beta}] = i\hbar \delta_{\alpha \beta}
\label{eq:CCRel}
\end{equation}
{}From the canonical commutators (\ref{eq:CCRel}) follows indeed the
quantization prescription
given below in terms of the basic variables $\xi$ of the problem,
\begin{equation}
\left \{ \xi_{A}, \xi_{B} \right \}_{PB} \longrightarrow \frac{1}{i\hbar}
\left [\hat{\xi}_{A}, \hat{\xi}_{B} \right ]
\label{eq:CCRxi}
\end{equation}
where quantal variables are distinguished by a hat, thus: $\hat{\xi}$.

The Hamiltonian H given in Eq. (\ref{eq:QCham}) can immediately be generalized
to an
operator Hamiltonian $\hat{H}$. Using the inverse relations (\ref{eq:CCxipqK}),
or more
explicitly the following relations,
\begin{equation}
\hat{\xi}_{2A} = \frac{4\lambda}{\sqrt{N}}\sum_{\alpha=1}^{\frac{1}{2}N-1}
D_{\alpha}
\hat{p}_{\alpha}\sin \varphi_{\alpha} \sin [(2A)\varphi_{\alpha} +
\frac{\pi}{4}]
 + \sqrt{\frac{2}{N}}K_{2}
\label{eq:opxie}
\end{equation}
and
\begin{equation}
\hat{\xi}_{2A-1} = \frac{4\lambda}{\sqrt{N}}\sum_{\alpha=1}^{\frac{1}{2}N-1}
C_{\alpha}
\hat{q}_{\alpha}\sin \varphi_{\alpha}\cos [(2A-1)\varphi_{\alpha} +
\frac{\pi}{4}]
 + \sqrt{\frac{2}{N}}K_{1}
\label{eq:opxio}
\end{equation}
one obtains the following expression,
\begin{equation}
\hat{H} = T(\hat{p}) + V(\hat{q})
\label{eq:TVpq}
\end{equation}
where
\begin{equation}
T(\hat{p}) = \sum_{A=1}^{\frac{N}{2}} \exp \left \{ \sqrt{\frac{2}{N}}K_{2} +
\frac{4\lambda}{\sqrt{N}}\sum_{\alpha=1}^{\frac{1}{2}N-1} D_{\alpha}
\hat{p}_{\alpha}
\sin \varphi_{\alpha}\sin [(2A)\varphi_{\alpha} + \frac{\pi}{4}] \right \}
\label{eq:opT}
\end{equation}
and
\begin{equation}
V(\hat{q}) = \sum_{A=1}^{\frac{N}{2}} \exp \left \{  \sqrt{\frac{2}{N}}K_{1} +
\frac{4\lambda}{\sqrt{N}}\sum_{\alpha=1}^{\frac{1}{2}N-1} C_{\alpha}
\hat{q}_{\alpha}
\sin \varphi_{\alpha}\cos [(2A-1)\varphi_{\alpha} + \frac{\pi}{4}]\right \}
\label{eq:opV}
\end{equation}
The Hamiltonian (\ref{eq:TVpq}) is expressed as a sum of terms depending
separately on the
momentum variables $\hat{p}$ and coordinate variables $\hat{q}$, respectively,
so there are no
essential quantum ordering problems in the quantization procedure outlined
above.

The quantization procedure considered by Volkov \cite{Volk} is not explicitly
canonical
and can therefore not be compared directly with the canonical standard
procedure given
above.

 \vfill\eject

\section{Summary and Discussion}

In this paper we have given a general analysis of the canonical structure of
the system of
differential equations which are known as the Lotka-Volterra model,
supplemented with
an antisymmetry condition (which is crucial for our analysis) which is also
frequently assumed
in connection with the Lotka-Volterra equations. Altogether this defines a
dynamical system,
with a finite number of degrees of freedom, which we call the Lotka-Volterra
system.

It has been shown that the Lotka-Volterra system always can be resolveded into
an explicitly
canonical Hamiltonian subsytem, involving an even number of canonical
equations, together with
a set of explicitly solvable constraints, which are time-dependent in general.
The set of
constraints is empty if the matrix which defines the interaction between the
various components
in the model, is regular. A set of canonical variables (pairs of
coordinates and momenta) is explicitly constructed for the Lotka-Volterra
system.

Furthermore, it has been shown that the Lotka-Volterra model gives rise to
several conserved
quantities (constants of motion) if the parameters of the model satisfy certain
conditions,
which have a general geometric characterisation as certain orthogonality
conditions. The
occurrence of such additional constants of motion has been termed a
"multi-Hamiltonian
structure".

As an illustration, a Lotka-Volterra system involving three degrees of freedom
has been
considered in detail.

An important special Lotka-Volterra system, which is associated with the
Liouville model
on a lattice has also been analysed in detail. The canonical structure and
Hamiltonian
formulation of the system in question has been shown to be a straightforward
consequence of the
general formalism developed in this paper. The canonical quantization for this
case, which
follows straightforwardly from the canonical structure of the underlying
classical model,
has been outlined in detail. This canonical quantization procedure differs in a
non-trivial
way from a previously proposed Hamiltonian procedure \cite {Fadtak,Volk} for
the model under
consideration.\\

{\bf Acknowledgements}\\

One of the authors (C.C) wishes to thank NORDITA for its hospitality and for
providing excellent
working conditions. Useful discussions with Professor Briitta Koskiaho are also
gratefully
acknowledged.

\vfill\eject

\end{document}